\documentclass[aps,prd,showpacs,amssymb,superscriptaddress, notitlepage,floats,floatfix,nofootinbib, twocolumn]{revtex4-1}
\usepackage{graphicx}
\usepackage{amssymb}
\usepackage{amsmath}
\usepackage{amsfonts}

\usepackage{bm}
\usepackage{mathrsfs}
\usepackage[colorlinks=true]{hyperref}
\usepackage[all]{hypcap}
\usepackage[utf8]{inputenc}
\usepackage{color}
\usepackage{multirow}
\usepackage{palatino}
\usepackage{float}
\usepackage{journals}
\usepackage{soul}
\usepackage{calrsfs}
\DeclareMathAlphabet{\pazocal}{OMS}{zplm}{m}{n}

\begin{document}
\title{The Inverse Problem for Hawking Radiation}
\author{Sebastian H. V\"olkel }
\email{sebastian.voelkel@uni-tuebingen.de}
\affiliation{Theoretical Astrophysics, IAAT, University of T\"ubingen, Germany}
\author{Roman Konoplya}
\email{roman.konoplya@gmail.com}
\affiliation{Institute of Physics and Research Centre of Theoretical Physics and Astrophysics, Faculty of Philosophy and Science, Silesian University in Opava, CZ-746 01 Opava, Czech Republic}
\affiliation{Peoples Friendship University of Russia (RUDN University), 6 Miklukho-Maklaya Street, Moscow 117198, Russian Federation}
\author{Kostas D. Kokkotas}
\affiliation{Theoretical Astrophysics, IAAT, University of T\"ubingen, Germany}
\date{\today}
\begin{abstract}
In this work we study the inverse problem related to the emission of Hawking radiation. We first show how the knowledge of greybody factors of different angular contributions $l$ can be used to constrain the width of the corresponding black hole perturbation potentials. Afterwards we provide a framework to recover the greybody factors from the actual energy emission spectrum, which has to be treated as sum over all multipole numbers. The underlying method for the reconstruction of the potential widths is based on the inversion of the Gamow formula, a parabolic expansion and the P\"oschl-Teller potential. We define a ``normalized'' energy emission spectrum that turns out to be very beneficial for the numerical fitting process, as well as for an improved qualitative understanding of how much information of the black hole potentials are actually imprinted in the spectrum. The connection to recent studies on the inverse problem using the quasi-normal spectra of ultra compact stars and exotic compact objects is discussed as well. In the appendix we show that the spectrum can be approximated surprisingly well and simply with a parabolic expansion of the peak of the classical black hole scattering potentials.
\end{abstract}
\maketitle
%
\section{Introduction}
%
Hawking radiation is one of the most important theoretical predictions in the application of quantum field theories to general relativity and alternative theories of gravity \cite{Hawking:1974sw}. Despite the current lack of experimental confirmation, it fostered almost countless theoretical works on the details of the emission process (see \cite{dur3052,Wald2001} and references therein). Among the most striking implications that followed, is the so-called information loss problem \cite{PhysRevLett.71.3743}, that arises if one applies the pillars of modern physical theories (quantum field theories and general relativity) to the quantum aspects of black holes.
\par
In this work we are interested in what we call the inverse problem. Assuming that the energy emission spectrum of Hawking radiation from a spherically symmetric black hole is provided, what can one learn about the black hole? It is well known that the emitted radiation is described by a black body being modified by greybody factors, which are related to the space-time of the black hole. In the following sections we outline a framework that can be applied to the greybody factors and to the entire energy emission spectra, to constrain the classical black hole perturbation potential, as well as to gain a more intuitive understanding of the individual contributions.
\par
The interest in the inverse problem for Hawking radiation is not a purely academic exercise, as it may seem from the first glance. The quantum corrections to the black hole metric, normally, are supposed to be negligibly small and unobservable for large astrophysical black holes, so that no information about quantum corrections should be expected from astrophysical observations of compact objects. On the contrary, behavior of miniature and primordial black holes, experiencing intensive Hawking radiation, will do strongly depend on the form of quantum corrections to gravity. Therefore, if Hawking radiation could be detectable in future experiments, this would allow us to trace back the geometry of the black hole and, at the end of the day, determine the form of quantum corrections.
\par
The inverse problem for the quasi-normal mode spectrum of spherically symmetric compact objects have been recently considered in a number of works. In \cite{paper2,paper5,paper6} it was shown that once the quasi-normal modes of ultra-compact stars or some types of wormholes are known, the effective potential can be reconstructed in an unique way, assuming general relativity and the validity of WKB theory. In \cite{Konoplya:2018ala} the reconstruction of the shape function from the high frequency (eikonal) quasi-normal spectrum for a broad class of wormholes was suggested. In contrast to the aforementioned studies, which are based on the quasi-normal mode spectrum, this work is based on the energy emission spectrum of Hawking radiation. Although its origin is quite different, the calculation of Hawking radiation involves the classical scattering problem related to finding the transmission coefficients, greybody factors,  through the black hole potential barrier. Within WKB theory and single barriers, this problem can be inverted to reconstruct a class of potentials that admit the given transmission. In contrast to the quasi-normal mode spectrum, Hawking radiation appears as a sum involving many greybody factors, which have to be recovered first. Despite the initially might expected challenges in the recovery of the individual greybody factors, we find that the problem can be split up in a multiple step approach which gives satisfactory results.
\par
In Sec. \ref{Hawking Radiation} we summarize the calculation of Hawking radiation being used in this work. The subsequent Sec. \ref{Inverse Problem} outlines the inverse problem methods. The applications and results of these methods are presented in Sec. \ref{Applications} and discussed in Sec. \ref{Discussion}. Our Conclusions can be found in Sec. \ref{Conclusion}. We provide additional material related to a simple and precise approximation of Hawking radiation in the appendix \ref{Appendix}. Throughout this work we use $G=c=\hbar=k_\text{B}=1$.
\section{Hawking Radiation}\label{Hawking Radiation}
In the following we give a short overview of the theoretical framework being used in this study.
\subsection{Energy Emission Spectra}
In this work we will assume that the black hole is in the state of thermal equilibrium with its environment in the following sense: the temperature of the black hole does not change between emissions of two consequent particles . This implies the canonical ensemble for the system. Therefore we work with the following description of the spectrum of Hawking radiation
\begin{align}\label{energy-emission-rate}
\frac{\text{d}E}{\text{d} t} = \sum_{l}^{} N_l \left| \pazocal{A}_l \right|^2 \frac{\omega}{\exp\left(\omega/T_\text{H}\right)-1} \frac{\text{d} \omega}{2 \pi},
\end{align}
were $T_H$ is the Hawking temperature, $A_l$ are the greybody factors, and $N_l$ are the multiplicities, which only depend on the space-time dimension and $l$. Details can be found in \cite{PhysRevD.80.084016} and references therein.
\subsection{Greybody Factors and Transmission}
It is well known that the greybody factors in the Hawking spectrum are related to the transmission through the black hole perturbation potentials with the corresponding spin of the field (see, for example, \cite{Page:1976df}). To obtain those, one has to solve the classical scattering problem of incoming radiation being transmitted or reflected at the potential barrier.
\par
The metric of a spherically symmetric black hole has the following form
\begin{equation}
\text{d} s^2 = -e^{2\nu} \text{d}t^2 + e^{2\lambda} \text{d}r^2 + r^2 (\text{d} \theta^2 +\sin^2 \theta \text{d} \phi^2).
\end{equation}
The linearized perturbation equations can be reduced to the following form
\begin{equation}\label{wavelike}
\frac{\text{d}^{2} \Psi}{\text{d} r_{*}^{2}} + (E - V(r))\Psi =
0,\,
\end{equation}
where the tortoise coordinate is defined as follows:
\begin{equation}
\text{d} r_{*}= e^{\lambda-\nu} \text{d}r.
\end{equation}
\par
We shall consider the wave equation (\ref{wavelike}) with the boundary conditions allowing for incoming waves from infinity. Owing to the symmetry of the scattering properties this corresponds to the scattering of a wave coming from the horizon. The scattering boundary conditions for (\ref{wavelike}) have the following form
\begin{equation}\label{BC}
\begin{array}{ccll}
    \Psi &=& e^{-i\omega r_*} + R e^{i\omega r_*},& r_* \rightarrow +\infty, \\
    \Psi &=& T e^{-i\omega r_*},& r_* \rightarrow -\infty, \\
\end{array}%
\end{equation}
where $R$ and $T$ are the reflection and transmission coefficients respectively.
\par
The effective potential has the form of the potential barrier which monotonically decreases at both infinities, so that the WKB approach \cite{PhysRevD.35.3621,PhysRevD.68.024018} can be applied for finding $R$ and $T$. Since the wave energy $E$ is real, the first order WKB values for $R$ and $T$ will be real \cite{PhysRevD.35.3621,PhysRevD.68.024018} and
\begin{equation}\label{1}
\left|T\right|^2 + \left|R\right|^2 = 1.
\end{equation}
Once the reflection coefficient is calculated, we can find the transmission coefficient for each multipole number $\ell$
\begin{equation}
\left|{\pazocal
A}_{\ell}\right|^2=1-\left|R_{\ell}\right|^2=\left|T_{\ell}\right|^2.
\end{equation}
Various methods for the computation of the transmission and reflection, which are energy dependent functions, exist in the literature. For quick and relatively accurate evaluation of the transmission and reflection coefficients for not small values of energy one can use the 6th order WKB formula \cite{PhysRevD.68.024018}. According to \cite{PhysRevD.68.024018} the reflection coefficient can be expressed as follows: 
\begin{equation}\label{moderate-omega-wkb}
R = (1 + e^{- 2 i \pi K})^{-\frac{1}{2}},
\end{equation}
where
\begin{equation}
K = i \frac{(\omega^2 - V_{0})}{\sqrt{-2 V_{0}^{\prime \prime}}} + \sum_{i=2}^{i=6} \Lambda_{i}.
\end{equation}
Here $V_0$ is the maximum of the effective potential, $V_{0}^{\prime \prime}$ is the second derivative of the
effective potential in its maximum with respect to the tortoise coordinate, and $\Lambda_i$  are higher order WKB corrections, which depend on up to $2i$th order derivatives of the effective potential at its maximum \cite{PhysRevD.68.024018}.
\par
For accurate calculations of the reflection/transmission coefficients at any energies we used two numerical methods. First, the shooting method, based on numerical integration from the event horizon up to the far region and consequent matching with the required asymptotic behavior \cite{Kokkotas:2010zd}. Second, a direct numerical integration of the wave equation from the transmission region backwards through the potential barrier to determine the transmission coefficients. This approach has been outlined and discussed in \cite{2011arXiv1101.2620R}. Both methods give similar results. From here and on, when mentioning \emph{exact transmissions} we will mean values obtained with the numerical approaches and not WKB theory.
\par
The semi-classical WKB treatment of the problem shows that the transmission for small energies below the barrier maximum can be approximated with the Gamow formula \cite{Gamow1928}.
\begin{align}\label{Gamow}
T(E)  = \exp\left(2 i \int_{x_0}^{x_1} \sqrt{E-V(x)} \text{d} x \right),
\end{align}
where $E$ is the energy, $V(x)$ the potential barrier, and $x_0$ and $x_1$ the classical turning points. In the application of black hole perturbation theory, the coordinate $x$ is the co-called tortoise coordinate $r^*$ and $E=\omega^2$. Higher order WKB descriptions for the transmission exist and have been applied in this context \cite{PhysRevD.82.084003}. In this work we will use eq. \eqref{Gamow} as basis for inverse problem. The reason for the use of this low order simplification is that the ``inversion'' of the Gamow formula is well known and defines a class of so-called width equivalent potentials. We discuss this in the following Sec. \ref{Inverse Problem}.
\section{Inverse Problem}\label{Inverse Problem}
In this section we outline the inverse method being used to reconstruct the black hole potential barrier widths for a given transmission, as well as the framework, if the energy emission spectra is provided.
\subsection{Individual Transmissions}
The inversion of the Gamow formula eq. \eqref{Gamow} has been derived and discussed in \cite{PhysRevA.18.1085,1980AmJPh..48..432L,2006AmJPh..74..638G}. From there it is known that providing the transmission $T(E)$ through a single potential barrier can not be used to uniquely determine its shape. In contrast, infinitely many so-called width equivalent potentials exist. A similar result exists for the Bohr-Sommerfeld rule of bound states in potential wells \cite{lieb2015studies,MR985100}. The width $\pazocal{L}$ of the potential barrier is given by
\begin{align}\label{width}
\pazocal{L} (E) \equiv x_1 - x_0 = \frac{1}{\pi} \int_{E}^{E_\text{max}} \frac{ \left(\text{d} T(E^\prime)/\text{d}E^\prime \right)}{T(E^\prime) \sqrt{E^\prime - E}} \text{d}E^\prime,
\end{align}
where $E_\text{max}$ is the maximum of the potential barrier. If it is not known, one can extrapolate it from where the transmission becomes $1/2$, which corresponds to the WKB result at the maximum of the barrier. To convert eq. \eqref{width} into a potential barrier one has to provide a function for one of the two turning points, $x_0(E)$ or $x_1(E)$, and invert the relation for $E$. Note that the provided turning point function has to exclude ``overhanging cliffs'' in the potential, which correspond to a multi-valued function for the potential. More details can be found in the aforementioned works \cite{PhysRevA.18.1085,1980AmJPh..48..432L,2006AmJPh..74..638G,lieb2015studies,MR985100}. The application of eq. \eqref{width} to individual transmission functions of black holes is shown in Sec. \ref{Applications}.
\par
Since the Gamow formula is not valid for energies around the peak of the barrier, we extend the reconstruction process. The P\"oschl-Teller potential is widely used in black hole theory to approximate calculations that involve the potential. In the form used here it is given by
\begin{align}
V(x) = \frac{V_0}{\cosh(ax)^2}.
\end{align}
It describes black hole potential barriers for energies around the maximum very well. We expand the potential barrier at the peak to parabolic order and uses an analytic formula for the transmission in the fitting procedure for energies around the peak. It is straight forward to identify the P\"oschl-Teller potential parameters with the two parabola parameters. This is used to approximate the peak region where the pure WKB treatment is not valid.
\subsection{Energy Emission Spectra}
The energy emission spectra is a sum over all angular contributions $l$, which maps the individual transmissions $T_l(E)$ in a non-trivial way in one function eq. \eqref{energy-emission-rate}. However, in order to apply the inverse method of Sec. \ref{Inverse Problem}, one has to find the individual transmission functions first. For this purpose we have worked out a framework that is based on some fairly general observations and explained in the following.
\subsubsection{Modeling Individual Transmissions}
The function describing the transmission through a potential barrier is by far not arbitrary. For single potential barriers studied in this work, it resembles a logistic function. We make use of this and parameterize any transmission with
\begin{align}\label{Transmission-model}
T_\text{fit, l}(E) = \frac{1}{1 + \exp\left( a_{l,0} + a_{l,1} E + a_{l,2}/\sqrt{E} \right)},
\end{align}
where $a_{l,i}$ are a priori unknown constants that have to be fitted to a given energy emission spectra. The index $l$ refers to the angular contribution. It is in principle straightforward to include higher order terms in cases where this ansatz does not yield sufficient precise results. As we show in the appendix \ref{Appendix}, the interesting contribution comes around the peak and is in this sense local. Higher oder terms in eq. \eqref{Transmission-model} could be used to capture long range effects of the potential, but those are very difficult to be reconstructed from the entire spectrum.
\subsubsection{Treatment of Energy Emission Spectra}\label{assumptions_specta}
The actual energy emission spectra will contain in principle infinitely many terms, but not all contribute in the same way. Making use of this observation simplifies the treatment of the full problem significantly. The weighting of each term with the $l$ dependent transmission. The individual transmissions act qualitatively like a high pass filter. For a given $l$, the threshold is located around the barrier maximum, which in the eikonal approximation is simply proportional to
\begin{align}
E_{\text{max}, l}  \sim l(l+1),
\end{align}
and therefore acts as a cut off for small $E$, but is transparent for large $E$. Thus, high $l$-terms only affect the spectra for large $E$, but can be neglected for small ones.
\subsubsection{Extract Transmissions}\label{Extract Transmissions}
\begin{figure}[h]
	\centering
	\includegraphics[width=8cm]{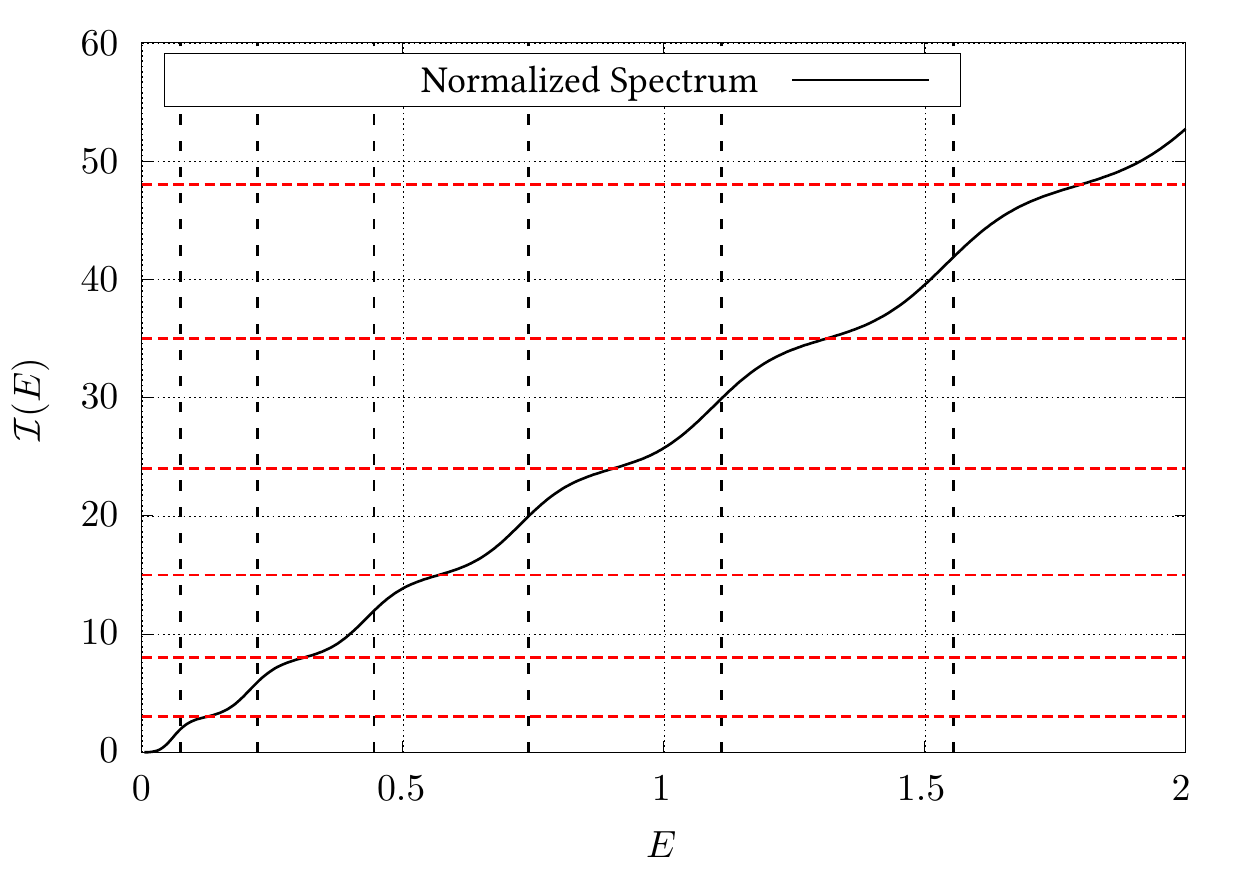}
	\caption{Here we show the normalized energy emission spectrum, defined in eq. \eqref{normalized-spectrum}, for the Schwarzschild case. The black dashed lines indicate the potential maximums of each potential barrier $E_{\text{max}, l}$ and match qualitatively with the saddle points. The red dashed lines show the summation of all multiplicities up to the $l-$th term $\sum_{i}^{l} (2i+1)$, starting from $l=1$ as lowest line. This line intersects the normalized spectrum in its flattest intervals between two consecutive potential maximums $E_{\text{max}, l}$.\label{norm_emission_spectrum}}
\end{figure}

Using the observations of the previous paragraph, it is straightforward to implement a numerical scheme that works in multiple steps. For a given transmission spectrum, one can start with the reconstruction of the lowest transmission function by fitting it to the set of parameters of our model eq. \eqref{Transmission-model}.  It is important to realize that this can only be done reasonably in an interval between neighboring potential maximums. The contribution of the transmission $T_l(E)$ to the spectrum dominates only in the interval $( E_{\text{max},l-1},  E_{\text{max},l})$. For smaller or larger energies it can be well approximated with $0$ or $1$, respectively. Once the parameters are determined in this range, we repeat with the fitting of the next transmission $T_{l+1}(E)$ in the subsequent interval. We find it useful to define a ``normalized'' spectrum, where we divide out all $l$ independent functions from the energy emission spectrum $\text{d}^2E/\text{d}w \text{d}t$, as described in eq. \eqref{energy-emission-rate}
\begin{align}\label{normalized-spectrum}
\pazocal{I}(E)
\equiv  \sum_{l}^{} N_l \left| \pazocal{A}_l \right|^2
\equiv \sum_{l}^{} I_l(E).
\end{align}
If the value of the Hawking temperature is not assumed to be known, it can be obtained from the following procedure. As it is shown in Fig. \ref{norm_emission_spectrum}, the normalized spectrum is well described with smoothened steps whose growth scales linearly with the energy. The case we show is for Schwarzschild, but it is similar for Reissner-Nordstr\"om, as we show in appendix \ref{Appendix_toy_model}. In contrast to this, the actually measured spectrum $I(E)$ falls off exponentially for large energies. The idea now is to find an approximation for the Hawking temperature by demanding that the normalized spectrum has to grow roughly linearly for large energies. In a second step we can read out the temperature by using the step structure. Going from the measured spectrum $I(E)$ to the normalized spectrum $\pazocal{I}(E)$ is done via
\begin{align}
\pazocal{I}_\text{rec}(E) = \frac{2 \pi}{\sqrt{E}} \left(\exp\left(\frac{\sqrt{E}}{T_\text{H, rec}} \right)-1 \right) \times I(E),
\end{align}
where $T_\text{H, rec}$ is the reconstructed Hawking temperature. The simple functional structure of the normalized spectrum allows one to precisely determine the Hawking temperature $T_\text{H, rec}$, as long as the perturbation potentials have the single barrier structure we assume in this work. The number of saddle points $N_\text{s}$ (counted from low to high energies) corresponds to the number of summed terms which are relevant up to the given energy. Contributions from higher terms at this energy value are being suppressed due to the transmission functions and therefore negligible. If one knows the full spectrum, one can determine $T_\text{H, rec}$ from demanding that the value of the measured spectrum has to match the sum of the multiplicities at the flattest point between $(E_{\text{max}, N_\text{s}}, E_{\text{max}, N_\text{s}+1})$. In Fig. \ref{norm_emission_spectrum} one can see that the flattest part of the normalized spectrum coincides very well with the summation of the multiplicities up to the $N_\text{s}$-th term. More details are being provided in the caption. 
\\~\\
Since this work assumes that the spectrum is known with high precision and no observational errors, this identification becomes in principle arbitrary precise by going to higher energies. Therefore we continue with the exact values. It is evident that observational errors on the measured spectrum would require a special treatment and put an error on the reconstructed temperature. However, the experimental access to Hawking radiation is currently not possible, which is the reason why we do not consider this limitation here.
\\~\\
Strictly speaking, when dealing with gravitational perturbations of static black holes in four dimensional spacetimes, two channels of perturbations come into play: the axial (vector) and polar (scalar) ones. They are represented by the corresponding vectors and scalars relatively the 2-sphere rotation group. In addition, the multiplicity factors are, in general case, different for different channels of gravitational perturbations. Therefore, when talking here about Hawking radiation of gravitons in the vicinity of the Reissner-Nordstr\"om black hole, we simply mean this single channel of perturbations which we considered here for purely illustrative purpose. The possibility of distinguishing different channels of gravitational perturbations was considered recently in \cite{Bhattacharyya:2017tyc} and we will not touch this problem here. It is evident that there is no such a problem for emission of particles of other spin.
\section{Applications and Results}\label{Applications}
In this section we demonstrate the results of the inverse method applied to the knowledge of different exact transmission functions as well as to several energy emission spectra. The transmissions and the energy emission spectra are obtained numerically by integration of the wave equation through the potential barrier for every given $E$ (see, for instance, \cite{Kokkotas:2010zd}). For the numerical implementation of the fitting we use the routines of CERN's ROOT Data Analysis Framework \cite{Brun:1997pa}. As examples to demonstrate the proposed methods, we study the vector case for the Schwarzschild and Reissner-Nordstr\"om black holes in four dimensional general relativity. The metric functions for these black holes are
\begin{align}
e^{2 \nu} = e^{-2\lambda} = 1 - \frac{2 M}{r} + \frac{Q^2}{r^2}
\end{align}
and the parameters of the black holes are provided in TABLE \ref{table}. We list the equations describing the two different types of perturbation potential in appendix \ref{appendix_potentials}.
\begin{table}[H]
	\centering
	\caption{Parameters of the studied black holes in this work.\label{table}}
	\begin{tabular}{|c||c|c|c|}
		\hline
		Model &		$M$	&	$Q$	 &	$T_\text{H}$ \\
		\hline
		Schwarzschild			&	1		&	0				&	$1/8\pi$ \\
		Reissner-Nordstr\"om	&	$2/3$	&	$1/\sqrt{3}$	&	$1/6\pi$ \\
		\hline
	\end{tabular}
\end{table}
\subsection{Reconstructing the Greybody Factors}
\begin{figure}[h]
	\centering
	\includegraphics[width=8cm]{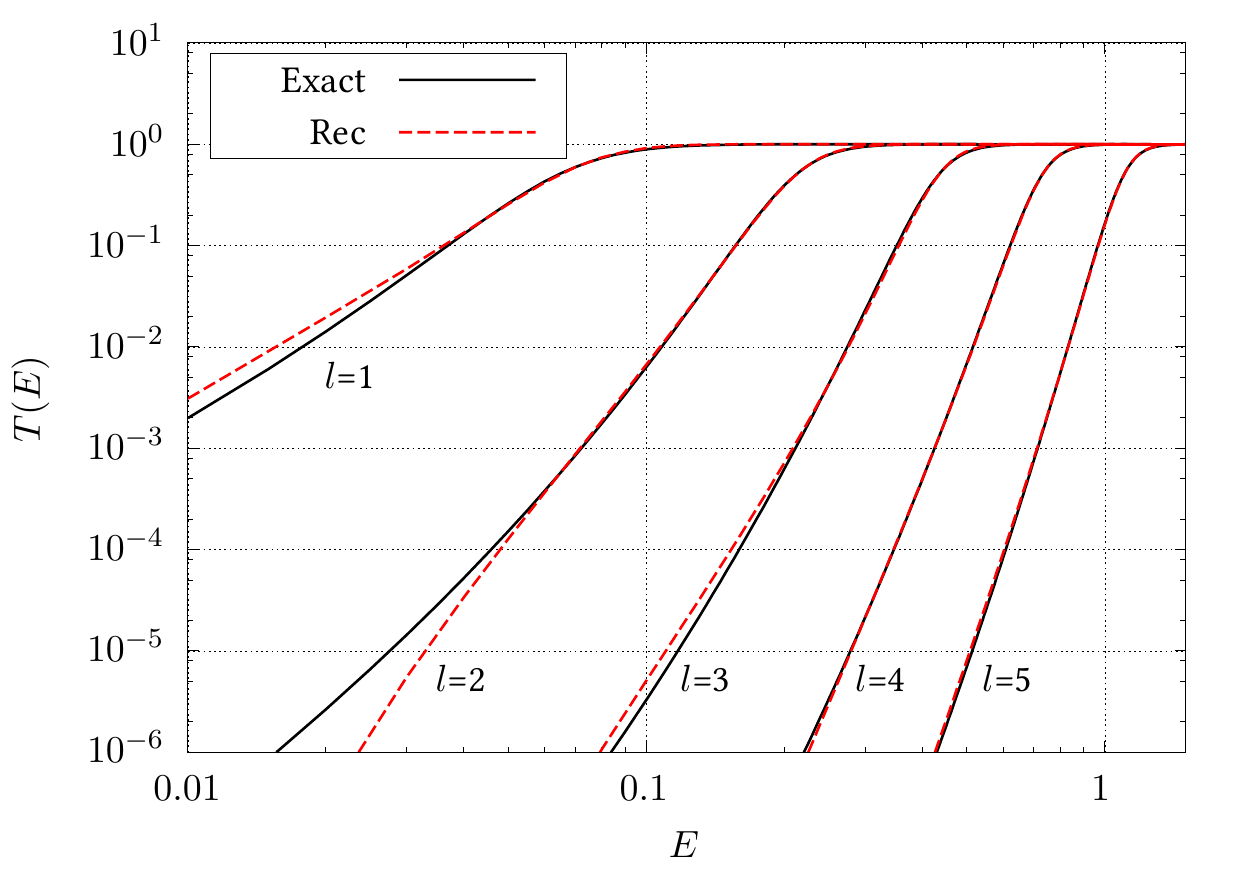}
	\caption{Reconstruction of the Schwarzschild transmissions $T_l(E)$ from spectrum fitting. \label{rec_transmission_SCH}}
\end{figure}

\begin{figure}[h]
	\centering
	\includegraphics[width=8cm]{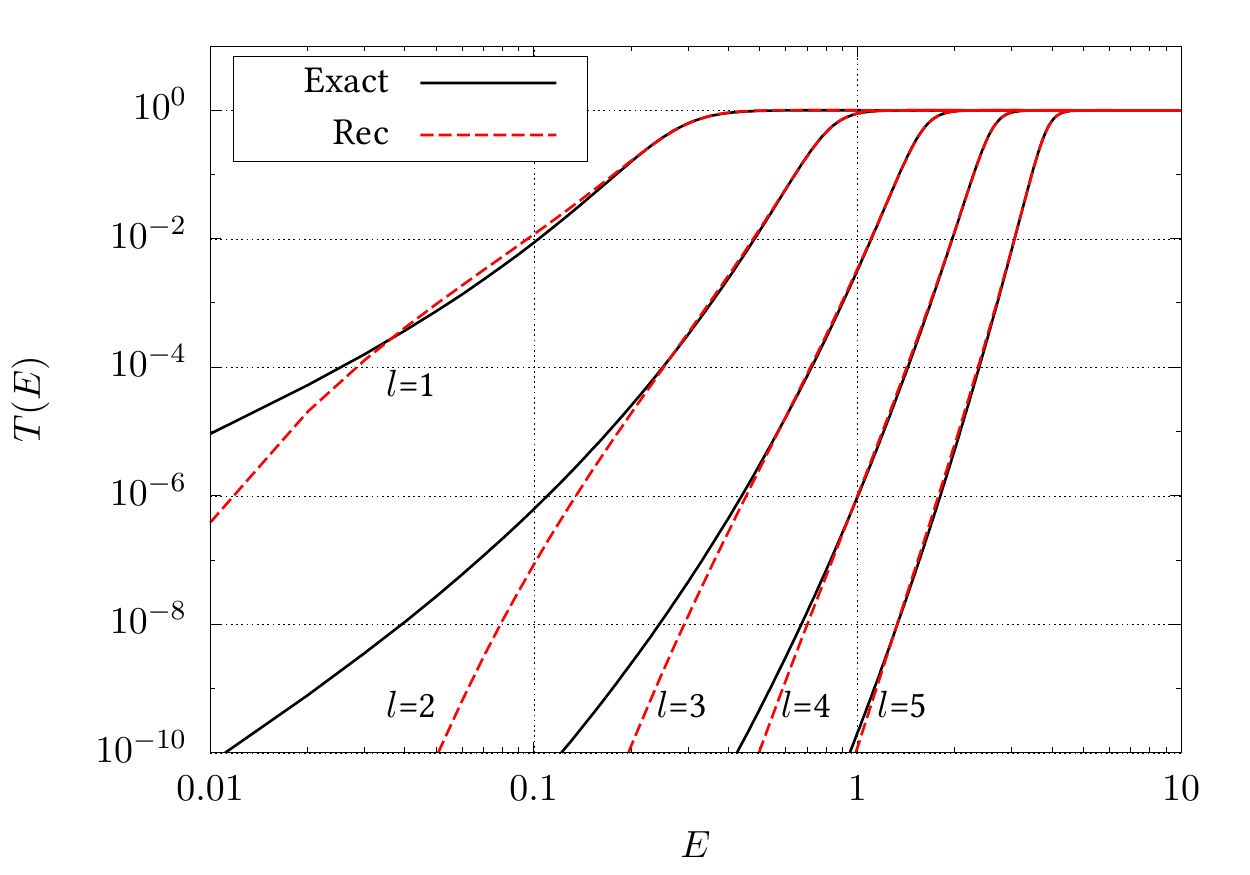}
	\caption{Reconstruction of the Reissner-Nordstr\"om transmissions $T_l(E)$ from spectrum fitting. \label{rec_transmission_RN}}
\end{figure}

By fitting numerically our expansion for the transmission to the normalized energy emission spectrum eq. \eqref{normalized-spectrum} we are able to reconstruct the transmissions for the first few $l$. Our results for this are shown in Fig. \ref{rec_transmission_SCH} for Schwarzschild and in Fig. \ref{rec_transmission_RN} for Reissner-Nordstr\"om. To obtain the normalized spectrum we used the correct value for the Hawking temperature, as explained in \ref{Extract Transmissions}.
\subsection{Reconstructing the Potential Widths}
\begin{figure}[h]
\centering
\includegraphics[width=8cm]{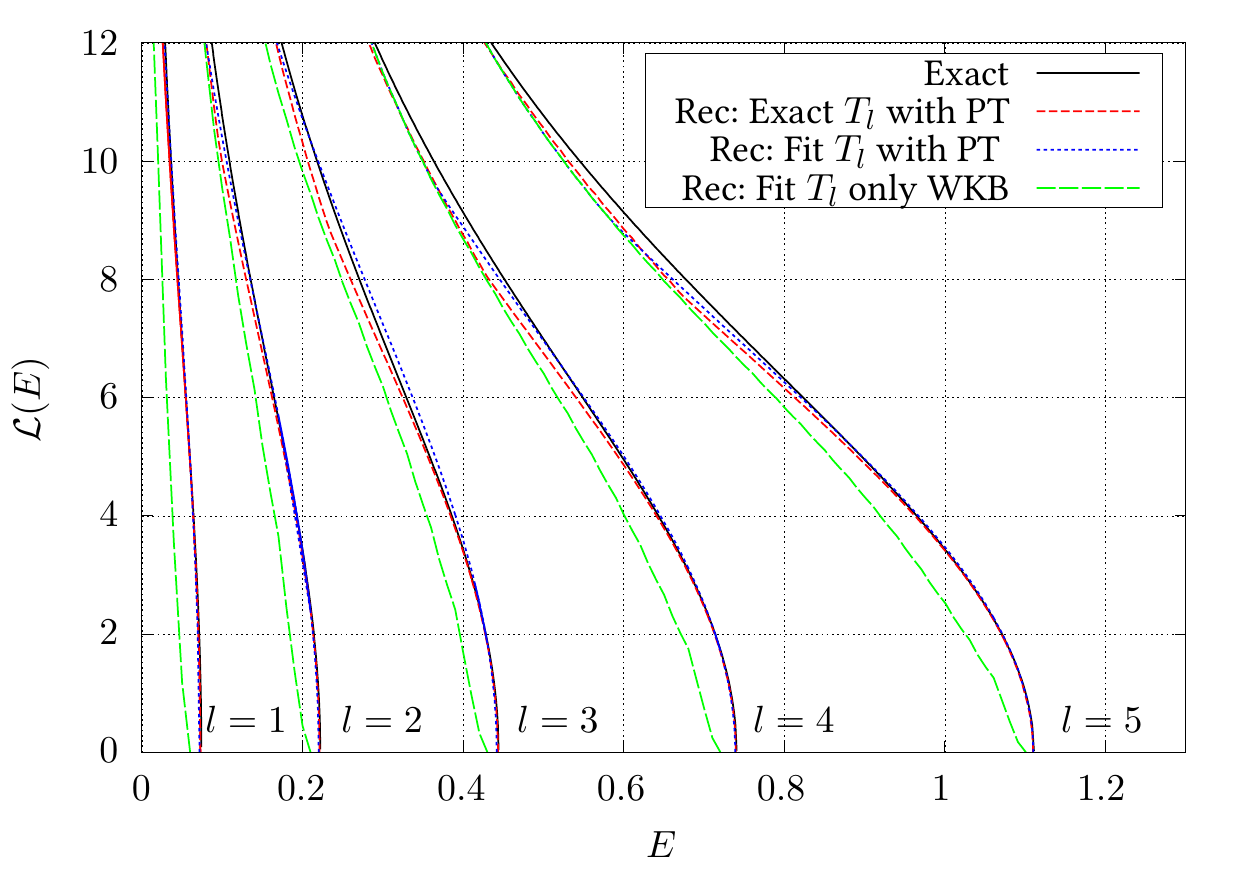}
\caption{Reconstruction of the Schwarzschild potential barrier widths $\pazocal{L}_l(E)$ from given transmissions $T_l(E)$. \label{rec_barrier_SCH}}
\end{figure}

\begin{figure}[h]
\centering
\includegraphics[width=8cm]{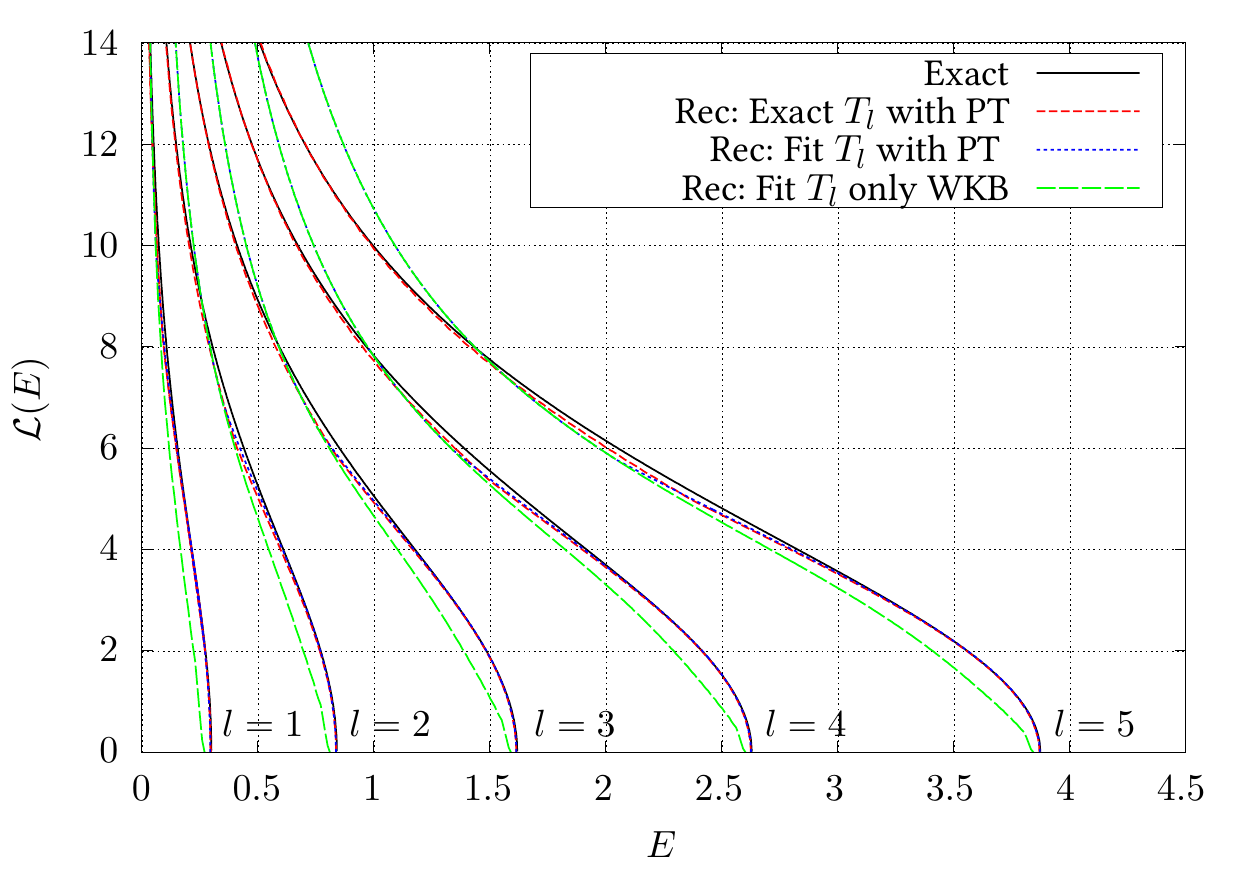}
\caption{Reconstruction of the Reissner-Nordstr\"om potential barrier widths $\pazocal{L}_l(E)$ from given transmissions $T_l(E)$. \label{rec_barrier_RN}}
\end{figure}

By using the reconstructed transmissions, we can now approximate the widths of the potential barriers. Our results are provided for Schwarzschild in Fig. \ref{rec_barrier_SCH} and for Reissner-Nordstr\"om in Fig. \ref{rec_barrier_RN}. Note that there are two aspects to be investigated. First, the general precision of the inverse method to obtain the potential widths from a given transmission. Second, how much the results for the inversion depend on the precision with which the transmission is provided. To address both questions we show the results using the numerically precisely calculated transmission (red dashed), the fitted transmission (blue dashed), and for further discussion the pure inverse WKB result (green dashed). Since the reconstructed transmissions $T_l$ are precise for energies from close to the potential maximum $E_{\text{max}, l}$ to the lower maximum $E_{\text{max}, l-1}$, it is not surprising that the exact and fitted transmissions yield comparable results there. However, we find clear deviations for smaller energies, which can be traced back to the imprecise reconstruction of the transmission in these regions, as shown in Fig. \ref{rec_transmission_SCH} and Fig.\ref{rec_transmission_RN}.
\section{Discussion}\label{Discussion}
In this section we discuss our findings and relate them to inverse spectrum methods that have recently been developed for ultra compact stars and exotic compact objects, such as gravastars and wormholes \cite{paper2,paper5,Konoplya:2018ala,paper6}.
\subsection{Reconstructing the Greybody Factors}
The individual contributions to the energy emission spectrum can be easily understood from the normalized energy emission spectrum. It intuitively demonstrates for which energies a given transmission encodes information of the potential barrier to the spectrum. We find that it only contributes on an interval around its barrier maximum. As a consequence one can only reconstruct the transmission in this interval precisely, but looses accuracy for smaller energies, where the contribution is exponentially small. In our examples the transmissions range over 10 orders of magnitude. We also find that due to the $l$ dependency of the multiplicities and the exponential suppression in the energy emission spectrum, higher $l$ contributions become negligible and therefore more challenging to reconstruct.
\subsection{Reconstructing the Potential Widths}
Our results for the reconstruction of potential barrier widths from the directly provided transmission functions of the Schwarzschild and Reissner-Nordstr\"om black holes show that the inverse method works very precise around the peak of the barrier and in the region far below. The method combines the parabolic transmission fit to find the P\"oschl-Teller potential width at the barrier maximum and merges it with the inverted Gamow formula for small energies. Here we make two comments. First, the precision at the peak region and for energies much below the peak is very good, taking into account the fairly simple structure of the method. Second, the least precise region is where the P\"oschl-Teller approximation becomes less valid, but the turning points are not too far away from each other. In case that the transmission have to be reconstructed from the spectrum, it should be expected that the reconstruction only works reliable in an interval between consecutive potential maximums, but not on the whole range.
\subsection{Connection to Related Inverse Spectrum Problems}
The here presented method for the reconstruction of transmission functions and black hole potential barrier widths has to be discussed in the context of other inverse methods that are based on gravitational wave observations. Most of them make use of the oscillation spectra of the objects \cite{1999LRR.....2....2K,1999CQGra..16R.159N,2009CQGra..26p3001B,2011RvMP...83..793K}, which can be for present matter or space-time itself. Relevant objects are neutron stars, black holes and more recently exotic compact objects. Besides the extensive numerical relativity based simulations for neutron stars and black holes, there are also multiple semi-analytic approaches. In the case of neutron stars, there are well established asteroseismology relations that can be used to constrain the compactness or equation of state of neutron stars by using different types of their fundamental oscillation modes \cite{1996PhRvL..77.4134A,Andersson:1997rn,1999MNRAS.310..797B,2019arXiv190110851M}. In addition to this, there are also spectral methods using masses and radii of neutron stars, to reconstruct their equation of state \cite{1992ApJ...398..569L,2012PhRvD..86h4003L,Lindblom:2013kra,Lindblom:2014sha,Lindblom:2018ntw}. Due to recent claims of tentative evidence of so-called echoes in the gravitational wave signals of merging black holes \cite{2017PhRvD..96h2004A,PhysRevD.98.044021,Abedi:2018pst}, there is a vast increase of interest in exotic compact objects. However, whether the challenging data analysis has been carried out sufficiently is under discussion \cite{2016arXiv161205625A,2018PhRvD..97l4037W,2018arXiv181104904N}. Among those rather exotic objects are toy models of ultra compact stars, gravastars, boson stars and wormholes. Although most of the models stretch today's well understood and tested physics far beyond trustworthy limits, they are in reach with future gravitational wave observations. For a current review and list of the extensive literature in this field we refer to \cite{Cardoso:2017njb}.
\par
Technically closer to the present work, it was recently shown how the knowledge of the quasi-normal mode spectra of ultra compact stars, gravastars and some type of wormholes can be used to reconstruct their perturbation potential and determine some of the underlying parameters \cite{paper2,paper5,paper6}. The methods presented there are based on the inversion of generalized Bohr-Sommerfeld rules and the Gamow formula. As it should be expected, the inversion is in general not unique. However, additional physical assumptions about the system can overcome this limitation. For more details of the method we refer to \cite{paper4}. In the present work, the uniqueness of the reconstructed perturbation potential is not given.
\par
In another work it was shown how the quasi-normal mode spectra of some wormholes can be used to reconstruct their shape function \cite{Konoplya:2018ala}. This method made use of higher order WKB methods that are well established in the field of black hole perturbation theory \cite{PhysRevD.35.3621,PhysRevD.68.024018}.
~\\
\section{Conclusions}\label{Conclusion}
In this work we have presented a semi-analytic method which uses the energy emission spectrum of Hawking radiation to reconstruct the greybody factors and from this the widths of the black hole perturbation potentials. By defining a ``normalized'' energy emission spectrum we were able to carry out a multi step fitting procedure to reconstruct the transmissions. The reconstruction is based on the numerical fitting of a suitable expansion of the classical transmission function and its direct identification with the greybody factors. In a second step we have combined the inversion of the WKB based Gamow formula with the analytic result for the transmission through a parabola around the peak, in order to reconstruct the potential width for small energies, as well as around the peak of the barrier. We outlined why higher $l$ terms are highly suppressed and thus do not contribute to the spectrum and are therefore not eligible for the inversion process.  After presenting the method and some general observations, we applied it to the vector case of the Schwarzschild and Reissner-Nordstr\"om black holes in four dimensions described by general relativity. In the appendix we show that the whole problem can be extremely simplified by noting that the energy emission spectrum can be very well described with a pure analytic parabolic model for the potential barriers, which might be a very useful approximation for analytic calculations in more complicated black hole potentials.
\acknowledgments
S. V. receives the PhD scholarship Landesgraduiertenf\"orderung. The authors acknowledge support from the COST Action GW-verse CA16104. R. K. would also like to thank the University of T\"ubingen for hospitality and the support of the grant 19-03950S of Czech Science Foundation (GACR). This publication has been prepared with the support of the ``RUDN University Program 5-100''.
%
%
\bibliography{literatur1}
%
\section{Appendix}\label{Appendix}
\subsection{Toy Model for the Energy Emission Spectrum}\label{Appendix_toy_model}
During this study on the inverse problem of Hawking radiation we noticed that only the region of the potential barrier around the maximum plays a dominant role in the energy emission spectrum. This simple observation can be used to work out a simple analytic toy model, for a quick and simple approximation of the spectrum. Here we present a model which is in the spirit of well known quasi-normal mode calculations for black holes. It is using a  parabolic expansion of the potential peak, as it is done in the Schutz-Will formula \cite{1985ApJ...291L..33S}. In the energy emission spectrum one sums over all $l$, but as being outlined in this work, not all contribute similar. For a given $T_l(E)$ one finds the following cases. Either it has negligible contribution, if $E \ll E_{\text{max},l}$; potential dependent contribution between 0 and 1, if $E_{\text{max},l-1} < E < E_{\text{max},l+1}$; becomes approximatively 1, if $E_{\text{max},l} \ll E$. We expect that the parabolic approximation for the transmission to be much more precise than the corresponding result in the quasi-normal mode application. Of course both approaches can only be valid in cases where the black hole potential can be represented by a single barrier.
\par
\begin{figure}[]
	\centering
	\includegraphics[width=8cm]{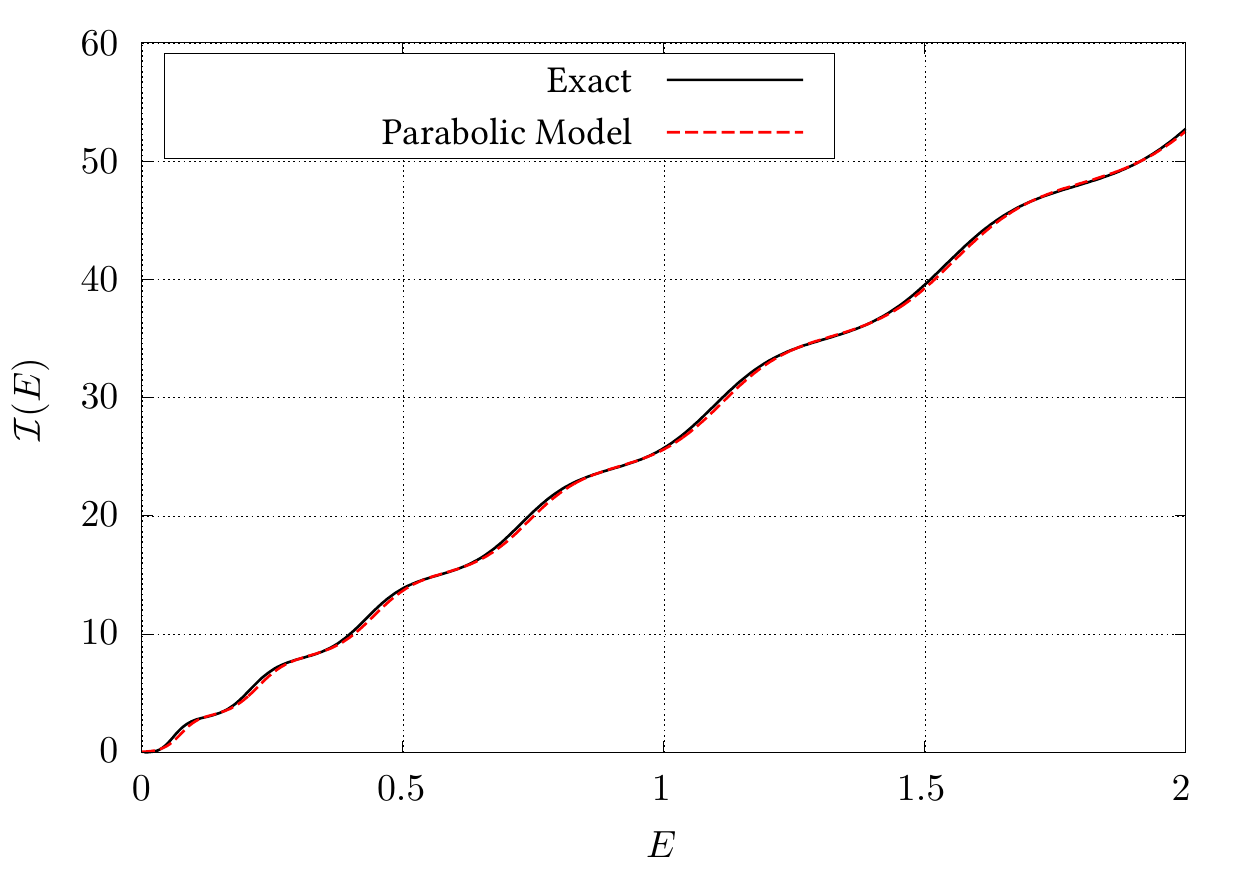}
	\caption{Comparison of the exact result (black solid) and parabolic approximation (red dashed) for the normalized energy emission spectrum of the Schwarzschild black hole. \label{approx_norm_emission_spectrum_SCH}}
\end{figure}
\begin{figure}[]
	\centering
	\includegraphics[width=8cm]{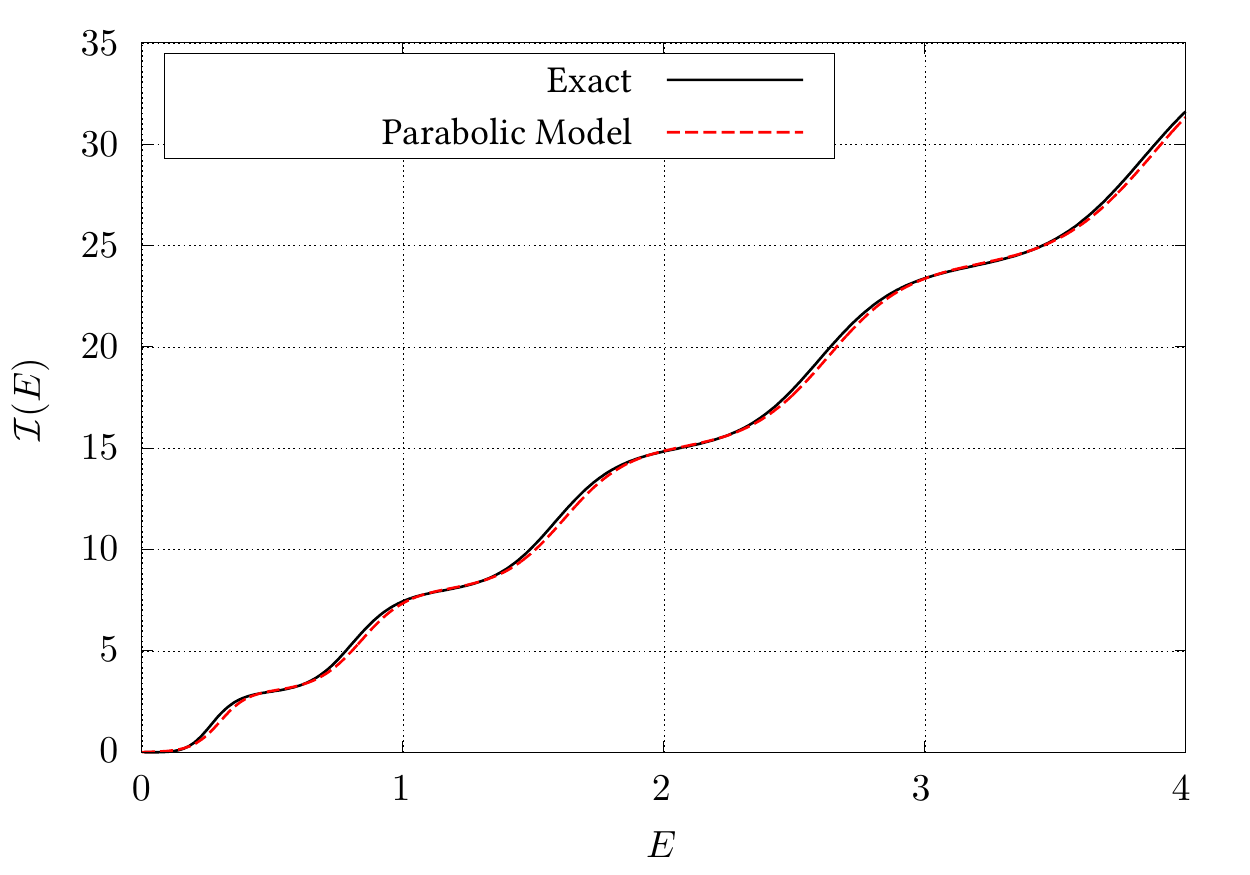}
	\caption{Comparison of the exact result (black solid) and parabolic approximation (red dashed) for the normalized energy emission spectrum of the Reissner-Nordstr\"om black hole. \label{approx_norm_emission_spectrum_RN}}
\end{figure}

The transmission through a parabolic potential barrier
\begin{align}
V(x) = V_{\text{max},l} - a_l x^2
\end{align}
described by WKB theory, and valid to describe the peak of the barrier, is given by
\begin{align}\label{model1}
T_l(E) = \left( 1+\exp\left(-\frac{\pi \left( E-V_{\text{max},l}\right)}{\sqrt{a_l}} \right)\right)^{-1},
\end{align}
see \cite{PhysRev.48.549,landau1981quantum} for $V_\text{max,l}=0$. The two parameters $V_{\text{max},l}$ and $a_l$ have to be matched with the black hole potential at the maximum $r^{*}_{\text{max},l}$. The identifications are
\begin{align}
V_{\text{max},l} \equiv V_\text{BH}(r^{*}_{\text{max},l}),
\qquad a_l \equiv - \frac{V^{\prime \prime }_{\text{max},l}}{2}.
\end{align}
The resulting normalized energy emission spectrum follows by using eq. \eqref{model1} as approximation for the greybody factors. We show the result for Schwarzschild in Fig. \ref{approx_norm_emission_spectrum_SCH} and Reissner-Nordstr\"om in Fig.  \ref{approx_norm_emission_spectrum_RN}. As one would naively expect from the Eikonal limit, one finds that the approximation becomes more and more precise the higher energies one considers.
\subsection{Perturbation Potentials}\label{appendix_potentials}
In this section we provide the equations describing perturbation potentials being used to calculate the transmission and reflection coefficients in this work.
\par
The perturbation potential describing a test electromagnetic field around the Schwarzschild black hole is given by
\begin{align}\label{eq-schwarzschild}
V(r) = \left(1-\frac{2M}{r} \right) \frac{l(l+1)}{r^2},
\end{align}
where the dependency on the tortoise coordinate is given implicitly as $r(r^*)$ and $l$ is the multipole number of the perturbation.
\par
The case of the Reissner-Nordstr\"om black hole is a bit more involved. In this work as a proof of principle we are only considering one of the four perturbations. The full derivation of the perturbation potential and further details can be found in \cite{doi:10.1098/rspa.1981.0142,1983mtbh.book.....C,Moncrief:1974gw,PhysRevD.10.1057,Moncrief:1975sb,PhysRevD.9.860,PhysRevD.37.3378}. The potential we use goes over to the pure vector perturbations of Schwarzschild eq. \eqref{eq-schwarzschild} in the limit of $Q=0$. This potential can be written as
\begin{align}
V(r) = \frac{\Delta}{r^5}\left(U + \frac{W \left( p_1-p_2\right)}{2} \right),
\end{align}
with the following abbreviations 
\begin{align}
\Delta &= r^2 -2Mr + Q^2,
\\
U &= \left(2nr+3M \right)W + \left(\bar{\omega}-nr-M \right) - \frac{2 n \Delta}{\bar{\omega}},
\\
p_{1,2} &= 3M \pm \sqrt{9M^2+8nQ^2},
\\
n &= \frac{(l-1)(l+2)}{2},
\\
W &= \frac{\Delta}{r \bar{\omega}^2}\left(2nr+3M \right) + \frac{nr+M}{\bar{\omega}},
\\
\bar{\omega} &= nr + 3M -\frac{2Q^2}{r}.
\end{align}
%
\end{document}